\documentclass[3p,times,procedia]{elsarticle}
\usepackage{nupha_ecrc}

\volume{00}

\firstpage{1}

\journalname{Nuclear Physics A}

\runauth{C. Shen and B. Schenke}

\jid{nupha}

\jnltitlelogo{Nuclear Physics A}

\usepackage{amssymb}

\usepackage[figuresright]{rotating}

\begin{document}

\begin{frontmatter}



\dochead{XXVIIth International Conference on Ultrarelativistic Nucleus-Nucleus Collisions\\ (Quark Matter 2018)}

\title{Dynamical initialization and hydrodynamic modeling of relativistic heavy-ion collisions}


\author{Chun Shen}
\author{Bj\"orn Schenke}

\address{Physics Department, Brookhaven National Laboratory, Upton, NY 11973, USA}

\begin{abstract}
We present a fully three-dimensional model providing initial conditions for energy and conserved charge density distributions in heavy ion collisions at RHIC Beam Energy Scan (BES) collision energies \cite{Shen:2017bsr,Shen:2017fnn}. The model includes the dynamical deceleration of participating nucleons or valence quarks. It provides a realistic estimation of the initial baryon stopping during the early stage of collisions. We also present the implementation of the model with 3+1 dimensional hydrodynamics, which involves the addition of source terms that deposit energy and net-baryon densities produced by the initial state model at proper times greater than the initial time for the hydrodynamic simulation. The importance of this dynamical initialization stage on hadronic flow observables at the RHIC BES is quantified.
\end{abstract}


\end{frontmatter}


\section{Introduction}
\label{Intro}

Over the past decade, significant progress has been made in our understanding of QCD matter properties in the extreme conditions created in relativistic heavy-ion collisions.
A wide range of collision energy, scanned in experimental heavy-ion programs, such as the Relativistic Heavy-Ion Collider (RHIC) Beam Energy Scan (BES) program and the NA61/SHINE experiment at the Super Proton Synchrotron (SPS), ensures extensive coverage in the temperature and baryon chemical potential of the produced nuclear matter. They provide us with a unique opportunity to explore the phase diagram of strongly interacting matter and to study the thermodynamic and transport properties of the quantum chromodynamics (QCD) matter in a baryon rich environment. These aspects are challenging to study from first principles lattice computations. The dynamical hybrid (viscous hydrodynamics + hadronic transport) model, on the other hand, plays an important role in elucidating the properties of the strongly-coupled Quark-Gluon Plasma (QGP) over a wide range of collision energies \cite{Heinz:2013th, Gale:2013da}. Moreover, the search for a critical point in the nuclear matter phase diagram relies crucially on a detailed understanding of the dynamics of collision systems.

Challenges arise at low collision energies $\mathcal{O}(10)$ GeV where the finite longitudinal extension of the colliding nuclei has to be taken into account. This leads to a substantial overlapping time $\tau_\mathrm{overlap} \sim 1-3$ fm/$c$ during which the two nuclei pass through each other. The pre-equilibrium dynamics during this time play an important role for understanding baryon stopping and density fluctuations along the longitudinal direction of collisions. Several models were developed to study heavy-ion collisions at low collision energies \cite{Cassing:2009vt,Karpenko:2015xea,Batyuk:2017sku,Denicol:2018wdp}.
Recently, dynamical initialization frameworks \cite{Shen:2017bsr,Okai:2017ofp,Akamatsu:2018olk,Kanakubo:2018vkl} were proposed to connect the pre-equilibrium stage of the system with hydrodynamics on a local collision-by-collision basis. In this framework, the hydrodynamic evolution starts locally at a minimal hydrodynamization time after the first nucleon-nucleon collision. The sequential collisions between nucleons that occur after will contribute as energy-momentum and net-baryon density sources to the hydrodynamic fields. 
In this proceeding, we provide some complementary results to Ref.~\cite{Shen:2017bsr,Shen:2017fnn}.

\section{Parameterizing the initial baryon stopping}

\begin{figure*}[h!]
  \centering
  \begin{tabular}{cc}
  \includegraphics[width=0.48\linewidth]{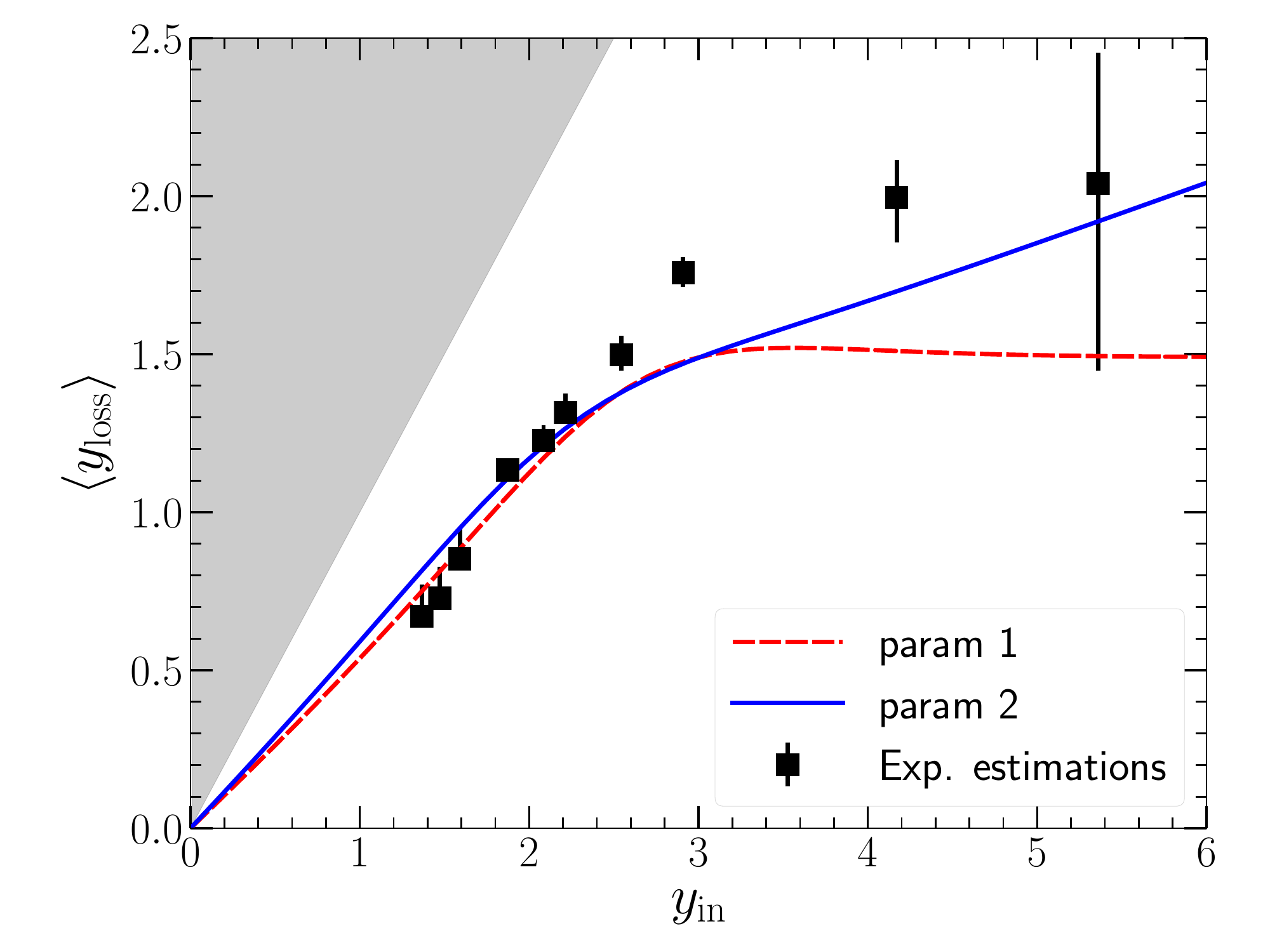} &
  \includegraphics[width=0.48\linewidth]{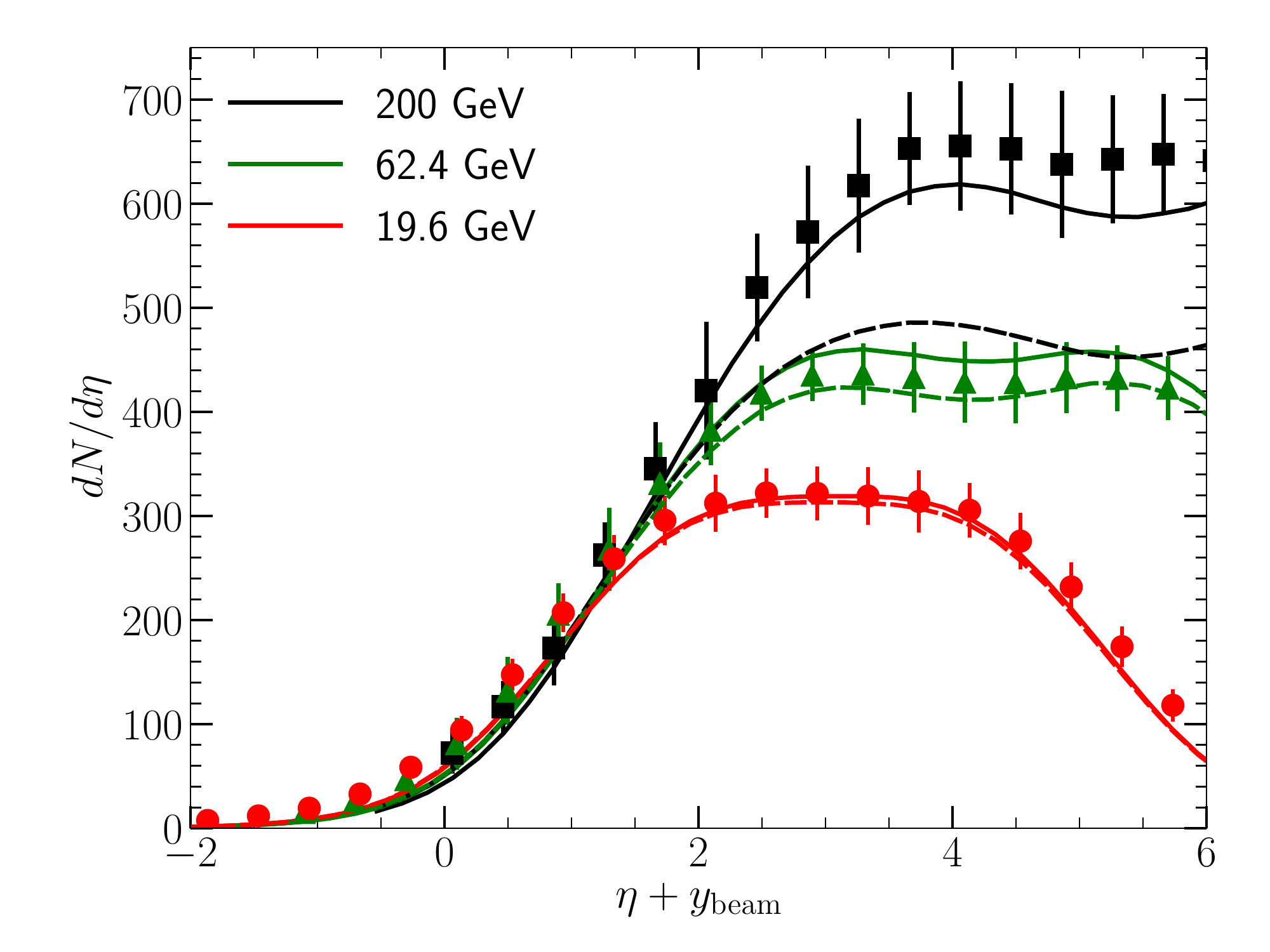}
  \end{tabular}
   \caption{Panel (a): Parameterizations of the rapidity loss for nucleon-nucleon collisions compared with experimental estimations \cite{Videbaek:1995mf,Blume:2007kw,Arsene:2009aa}. Panel (b): Pseudorapidity distributions of the charged hadron multiplicity at different collision energies shifted by their corresponding beam rapidities. Model calculations are compared with the PHOBOS measurements \cite{Back:2005hs}. Dashed lines are results using the parameterization 1 and solid lines are for parameterization 2.}
  \label{fig1}
\end{figure*}

Baryon stopping during the initial collision stage is hard to calculate from first principles. In our framework, a classical string deceleration model is employed \cite{Shen:2017bsr, Bialas:2016epd} which allows us to extract the baryon stopping information from phenomenological comparison with experimental measurements. A pair of colliding nucleons pulls out a string from each other and gets decelerated according to,
\begin{equation}
\frac{dE}{dz} = -\sigma \qquad \mbox{and} \qquad \frac{d p_z}{dt} = - \sigma,
\end{equation}
where $\sigma$ is the string tension constant. Particles' energy and momentum lost during the deceleration period is deposited uniformly along the string to form the hot medium in the central rapidity region. The remaining energy, momentum, and the baryon number of the collided nucleons are fed into the hydrodynamic fields at the string endpoints. In such setup, we conserve energy, momentum, and net baryon number. 
We assume that the rapidity loss during the deceleration period depends on the nucleon's initial rapidity in the pair rest frame. Fig.~\ref{fig1}a shows the two parameterizations of the rapidity loss functions we tested in this set of simulations. The largest difference between the two is at large incoming rapidities, $y_\mathrm{in} > 3$.
We also compare with the experimental estimations for baryon stopping \cite{Videbaek:1995mf,Blume:2007kw,Arsene:2009aa}. The string deceleration time is fixed to be $\Delta \tau = 0.5$ fm/$c$. 
%
%
Fig.~\ref{fig1}b shows the pseudo-rapidity distributions of the charged hadron multiplicity for 0-5\% Au+Au collisions at three different collision energies. Parameterization I (dashed lines) results in a smaller charged hadron multiplicity in the central rapidity region at 200 GeV because it underestimates the energy loss for nucleons with large initial rapidity. Parameterization II provides a reasonable description of the experimental data at all three collision energies.  Hence, the collision energy dependence of the charged hadron multiplicity is sensitive to how we parameterize the energy loss (or baryon stopping) in the model. Furthermore, shifting the distribution by the beam rapidity demonstrates that particle production near the fragmentation region is universal between different collision energies in both our model and the data. 

\section{Dynamical initialization vs instantaneous hydrodynamization}

\begin{figure*}[t!]
  \centering
  \begin{tabular}{cc}
  \includegraphics[width=0.47\linewidth]{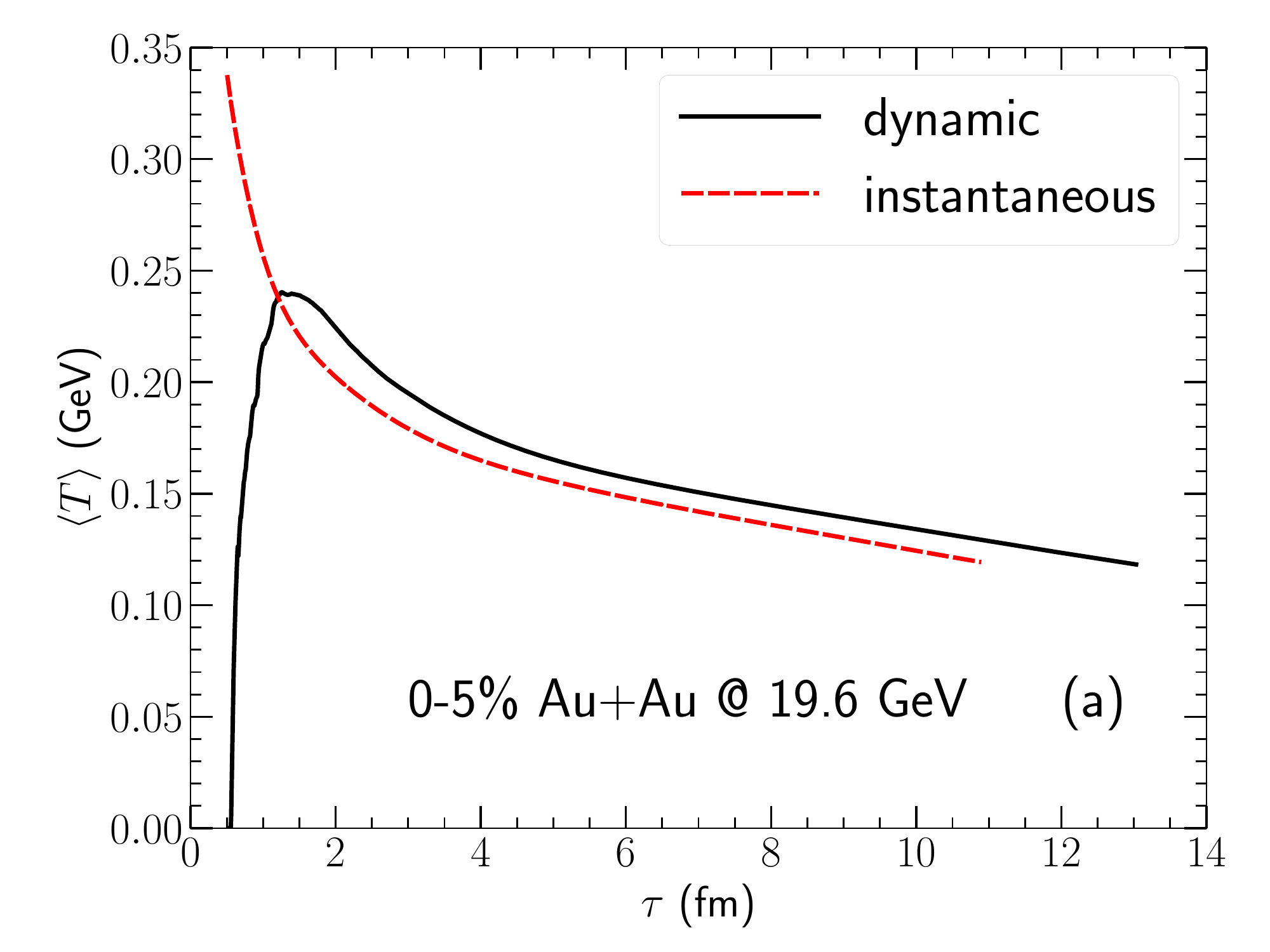} &
  \includegraphics[width=0.47\linewidth]{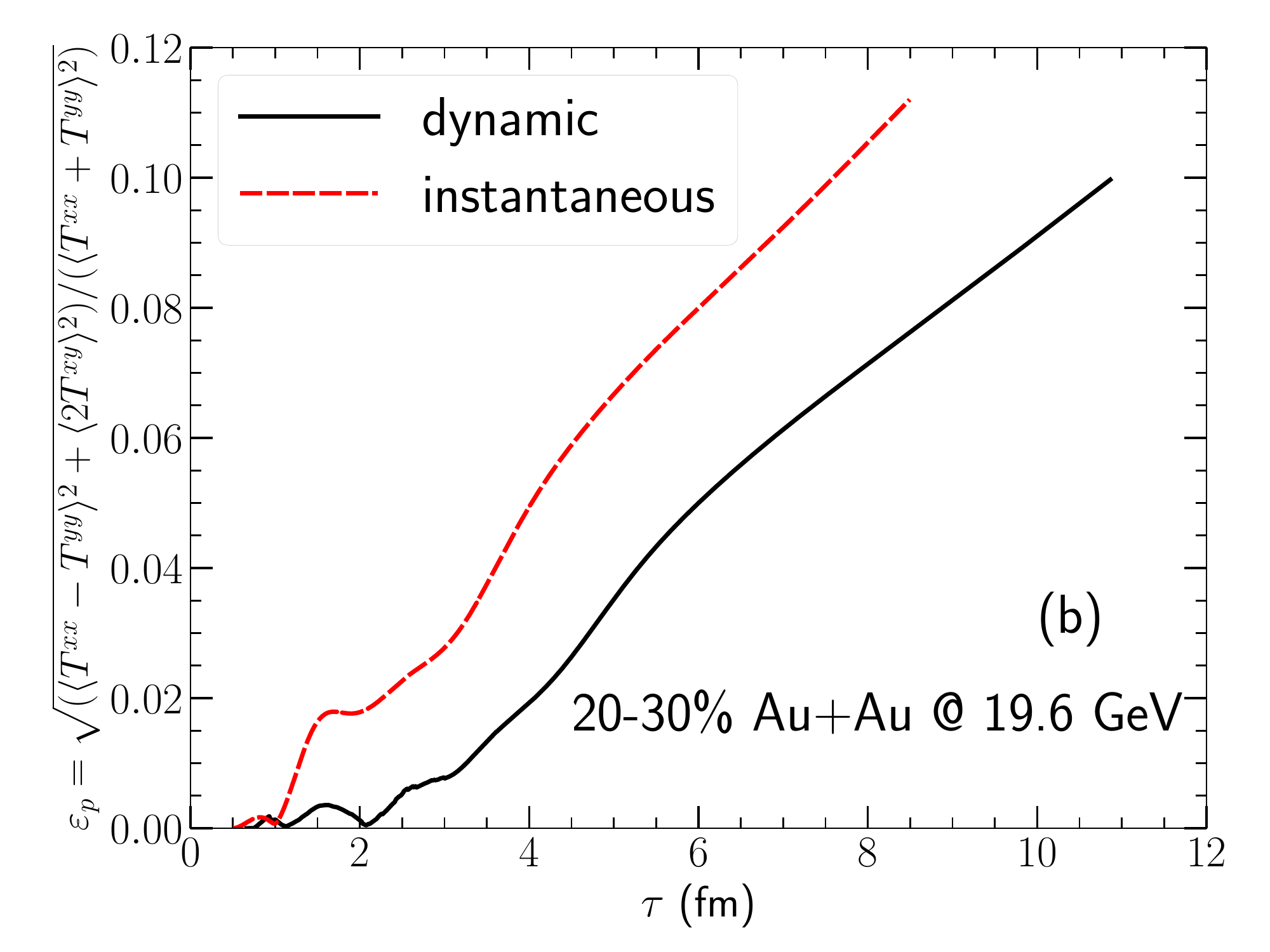}
  \end{tabular}
   \caption{Time evolution of the average fireball temperature (a) and system's momentum anisotropy (b) comparing dynamical initialization and instantaneous hydrodynamization scenarios.}
  \label{fig2}
\end{figure*}

In Fig.~\ref{fig2}, we explore the importance of the dynamical initialization for hydrodynamic evolution. We choose to study a central Au+Au collision at 19.6 GeV in which the colliding nuclei take about $\tau_\mathrm{overlap} = 1.5$ fm/$c$ to pass through each other. In the dynamical initialization case, the energy-momentum sources start to be fed into the hydrodynamic fields at $\tau_0 = 0.5$ fm/$c$. Individual nucleons inside the nucleus go through multiple collisions and deposit energy and momentum into the medium. In contrast, to set up the instantaneous hydrodynamization for the same collision event, we deposit all the energy-momentum sources at a fixed proper time $\tau_0 = 0.5$ fm/$c$.
%
%
In Fig.~\ref{fig2}a, the evolution of the system's averaged temperature during the first 2 fm/$c$ illustrates the difference between the two scenarios. The dynamical initialization heats up the system gradually as source terms are deposited into the medium along the evolution. Compared to the case of instantaneous hydrodynamization, the fireball also lives about 2 fm/$c$ longer. The development of the system's momentum anisotropy, $\varepsilon_p $, for the two cases is compared in Fig.\ref{fig2}b. Both the temperature and momentum anisotropy evolutions reflect that the transverse dynamics is developed more slowly in the dynamical initialization setup. Such a difference in hydrodynamic evolution results in about 5-10\% reduction of the produced particle mean-$p_T$ and elliptic flow coefficient. More importantly, because most of the difference happens during the early stage of the evolution, electromagnetic probes, such as direct photons, are more sensitive to the dynamical initialization \cite{Paquet:QM2018}.

\begin{figure*}[h!]
  \centering
  \begin{tabular}{cc}
  \includegraphics[width=0.47\linewidth]{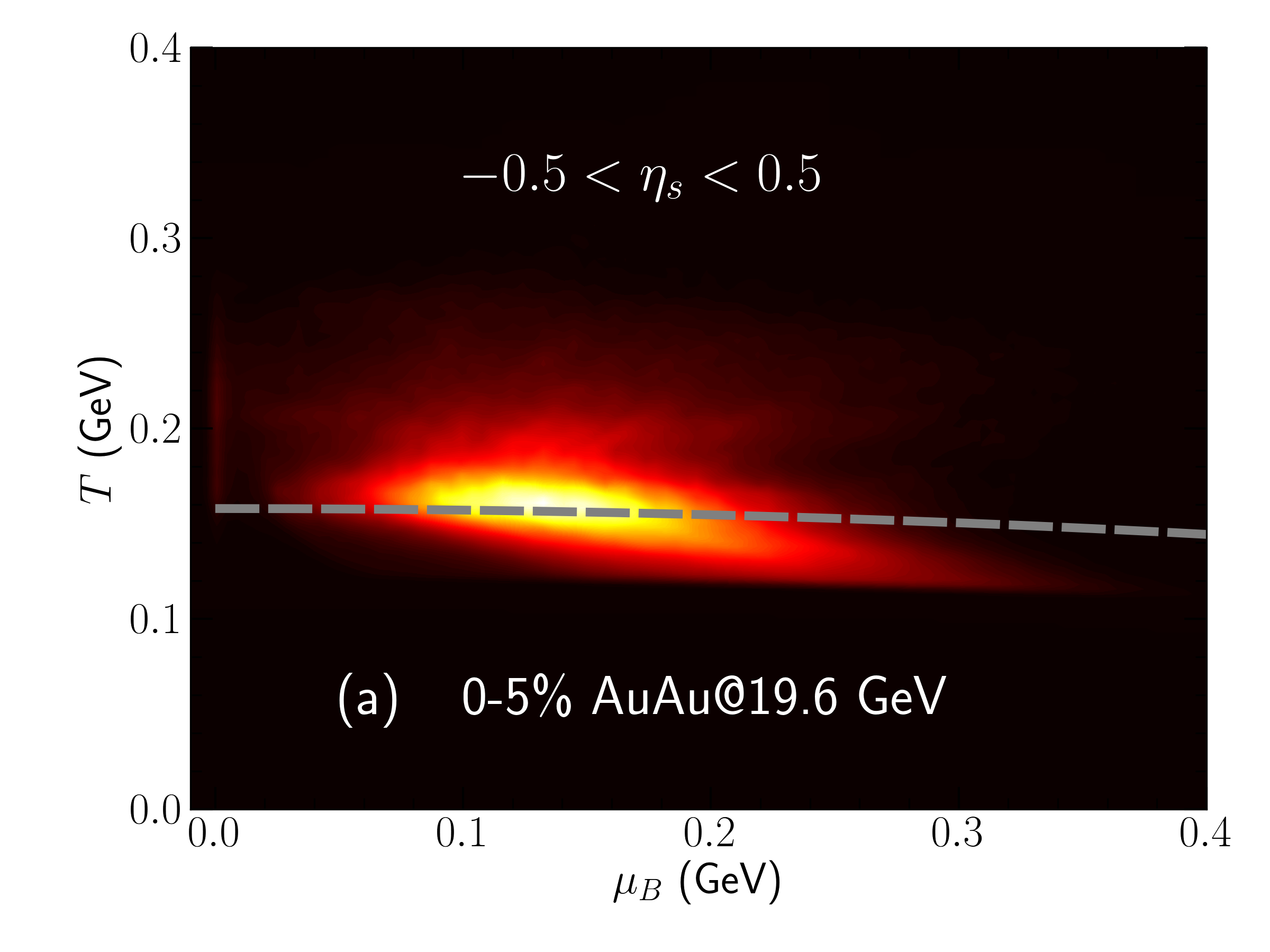} &
  \includegraphics[width=0.47\linewidth]{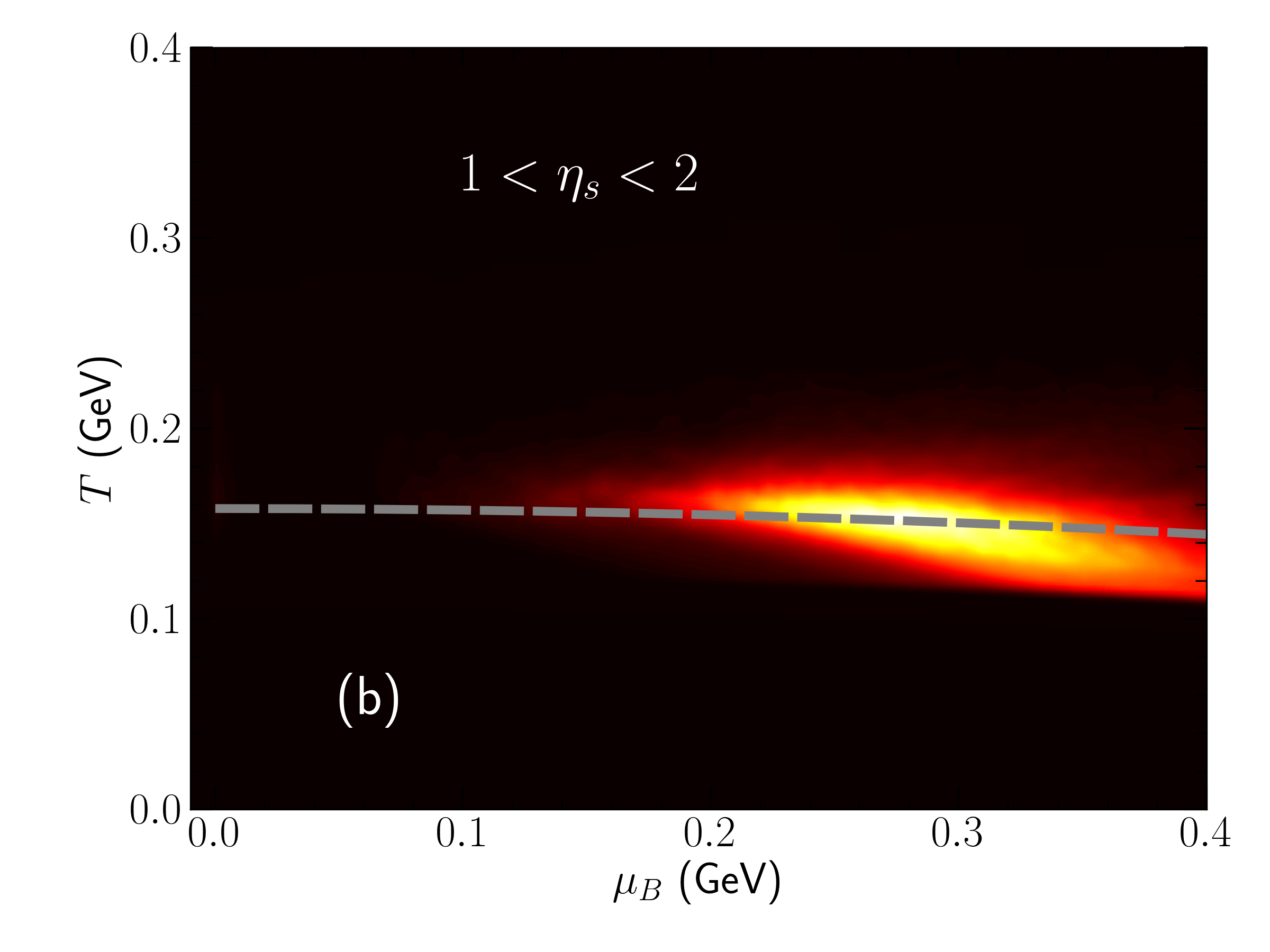}
  \end{tabular}
   \caption{The time accumulated fireball trajectories on the $T-\mu_B$ phase diagram for one 0-5\% Au+Au collisions at 19.6 GeV in central (a) and forward (b) rapidity regions. The grey dashed curve indicates the chemical freeze-out line from the STAR collaboration \cite{Adamczyk:2017iwn}. The brightness is proportional to the local energy density weighted space-time volume of the fireball.}
  \label{fig3}
\end{figure*}
%

\section{Hydrodynamic mapping of heavy-ion collisions to the QCD phase diagram}

Our dynamical framework can provide detailed information about the evolution of the bulk medium.  
The fireball trajectories and how fast every fluid cell travels through the phase diagram are essential information for modelling how critical fluctuations evolve out-of-equilibrium as the fireball passes the critical region. Such a study with event-by-event heavy-ion collisions was first performed in Ref. \cite{Bass:2012gy}.
The trajectory of one typical 0-5\% Au+Au collision at 19.6 GeV is shown in Fig.~\ref{fig3} in the central and forward rapidity regions. Because of the initial state fluctuations, the trajectory is not well confined as a thin line in the $T-\mu_B$ plane. A variance of $\sigma_{\mu_B} \sim 60$ MeV in net baryon chemical potential is observed as the fireball crosses the chemical freeze-out line in the transition region. This posts a challenge to precisely pinpoint the location of the critical point in the $T-\mu_B$ phase diagram.
The fireball trajectory in a forward rapidity region in Fig.~\ref{fig3}b scans through a larger $\mu_B$ but a lower $T$ region. The rapidity dependence of experimental observables can provide an additional dial to explore the $T$ and $\mu_B$ dependence of nuclear matter properties complementary to the variation in collision energy \cite{Brewer:2018abr}.

\medskip
\noindent {\bf Acknowledgements}
This work is supported in part by the U.S. Department of Energy (DOE), Office of Science, Office of Nuclear Physics, under Contract No. DE-SC0012704, and within the framework of the Beam Energy Scan Theory (BEST) Topical Collaboration. This research used resources of the National Energy Research Scientific Computing Center, which is supported by the DOE Office of Science under Contract No. DE-AC02-05CH11231.





\bibliographystyle{elsarticle-num}
\bibliography{references}







\end{document}